\documentclass[a4paper,11pt]{article}
\pdfoutput=1 % if your are submitting a pdflatex (i.e. if you have
\usepackage{jheppub} % for details on the use of the package, please
\usepackage{braket}
\usepackage[dvipsnames]{xcolor}
\usepackage{float}
\usepackage{caption}
\usepackage{subcaption}

\author[a]{David Berenstein,}
\author[a]{Elliot Maderazo,}
\author[a]{Robinson Mancilla}
\author[b,c]{and Anayeli Ram\'irez}

\affiliation[a]{Department of Physics, University of California at Santa Barbara,\\California 93106, USA}
\affiliation[b]{Dipartimento di Fisica, Universit\`a di Milano--Bicocca,\\Piazza della Scienza 3, I-20126 Milano, Italy}
\affiliation[c]{INFN, sezione di Milano--Bicocca}

\emailAdd{dberens@physics.ucsb.edu, emaderazo@ucsb.edu, rhmancilla@ucsb.edu, Anayeli.Ramirez@mib.infn.it}

\usepackage{braket}
\usepackage[dvipsnames]{xcolor}
\usepackage{float}
\usepackage{caption}
\usepackage{subcaption}

\title{Chaotic  LLM billiards.}

\abstract{We study null geodesics of the ten-dimensional LLM geometries. In particular, we show that there are a subset of these null geodesics that are confined to the LLM plane. 
The effective dynamics of these in-plane geodesics is that of a Hamiltonian system with two degrees of freedom (a phase space of dimension 4). We show that these are 
chaotic. In the two-coloring of the LLM plane, if they start in the empty region, they cannot penetrate the filled region and viceversa. The dynamical problem is therefore very similar to that 
of a billiards problem with fixed obstacles.  We study to what extent  LLM geometries with many droplets may be treated as an incipient black hole and draw analogies with the fuzzball proposal.
 We argue that for in-plane null geodesics deep in the interior of a region with a lot of droplets, in order to exit towards the $AdS$ boundary they will need to 
undergo a process that resembles diffusion.  This mechanism can account for signals getting lost in the putative black hole for a very long time.}

\begin{document} 
\maketitle
\flushbottom
\section{Introduction}
\label{sec:intro}

What is a black hole? In the classical theory of general relativity, the notion of a black hole has a precise definition. Colloquially, it is the region of spacetime situated behind a horizon from which no information can escape. Usually, there is also a singularity in the future of any infalling observer. 

Once we include quantum mechanics, especially in view of the gauge/gravity duality \cite{Maldacena:1997re}, describing the notion of a black hole is harder. Noting that semi-classical stationary black holes have a temperature \cite{Hawking:1975vcx}, one can associate black holes with a particular class of thermal states. This serves us well in equilibrium. Outside of equilibrium, however, it is very hard to extract in general the notion of a horizon from a quantum field theory dual computation. Part of the reason is that the notion of a horizon in general relativity is teleological. It is tied to future infinity: the horizon for an $AdS$ black hole is the surface that divides the events that a future asymptotic observer on the boundary of $AdS$ space might be able to see, versus the ones that are invisible to them. The horizon might also depend on how we choose to act on the quantum system on the boundary later on. This requires a notion of time in the bulk. We are thinking here of the spacetime being foliated by Cauchy slices labelled by a parameter that we choose to call the time variable. We want that time variable to coincide with the time variable on the boundary of $AdS$. So long as we can make different choices later on, a portion of spacetime at an instant of time might be situated behind the horizon or not, depending on some event that has not happened yet. In that sense, the field theory dual of a classical horizon is not an observable.

 Nevertheless, the presence or absence of a horizon can be used as an order parameter distinguishing two phases of matter: confinement and deconfinement \cite{Witten:1998zw}. This is simplest in a static setting with large black holes where one can Wick rotate the theory to a Euclidean geometry. The reason to choose large black holes is that they are thermodynamically stable \cite{Hawking:1982dh}. 
 Namely, they have positive specific heat. The Hawking-Page first order phase transition is identified with the confinement/deconfinement transition.
 The deconfined phase, in this case, has a different topology of the Wick rotated Euclidean spacetime than the confined phase. It should be noted that the same Euclidean topology is assigned to small black holes in $AdS$ space. These are black holes that behave like Schwarzschild black holes in flat space: they have negative specific heat. Reasoning by similarity, small black holes have been argued to be a partially deconfined phase \cite{Berenstein:2018lrm,Hanada:2018zxn} (see also the earlier \cite{Asplund:2008xd}), where only a subset of the color gauge group is deconfined. Their entropy is of order $\epsilon N^2$, with $\epsilon\lesssim 1$. Usually, $\epsilon$ is very small but finite, so there is not enough energy in the system to excite most of the color degrees of freedom (this would require an energy strictly larger than an order one constant times $N^2$ in free field theory). To solve the problem of the thermodynamic instability of these black holes one needs to work in the microcanonical ensemble: one fixes the mass, not the temperature. Since the horizon can become arbitrarily small, when these black holes are sufficiently small they can evaporate rapidly into the bulk. The question of whether there is a black hole or not becomes harder to analyze. There is no clear dividing line in the quantum gravity theory between spacetimes with black holes and spacetimes without them, essentially because black holes can evaporate.

 More to the point, there are many examples of geometric solutions of Einstein's equations that are horizon free, stationary and have the same asymptotic behavior as a black hole with the same quantum numbers. These solutions have bubbles (they have different topologies with many non-contractible cycles with fluxes through them) and many times only start to differ significantly from a corresponding black hole solution near the region where the horizon would have been located. These solutions usually arise from states that satisfy a supersymmetric bound between their charges and the mass (they are BPS states). The corresponding black hole is extremal and has zero temperature.  The fuzzball proposal \cite{Mathur:2005zp} (see also \cite{Bena:2022rna} for an overview of these ideas)   states that the black hole is just a coarse-grained version of these possible microstate geometries. Moreover,  the horizon appears due to our lack of knowledge of what is occurring on the {\em inside region}, broadly speaking. The proposal states that  ``horizons and singularities are artifacts of attempting to describe gravity using a theory that has too few degrees of freedom to resolve the physics" \cite{Bena:2022ldq}.
 
 More precisely, these extremal black hole solutions have a well defined throat region and it is in this throat region where the differences between the different geometries appear. In this case, the right question to ask is how can any of the typical microstate geometries seem to us as if it were a black hole? 

That question is addressed by throwing observers (matter) into the black hole and asking what they see (seeing what comes out). The answer to the question becomes phenomenological: can an individual microstate behave qualitatively similar to a black hole? For example, do signals generally take a long time to get out? Are there large tidal forces that rip objects apart? 
If such tidal forces exist, the fuzzballs would appear to destroy information to an outside observer.
These questions are computable: given details of the microstate geometry, one can throw matter inside and follow its motion in the interior. One can then wait until the matter comes out again.

Our goal in this paper is to understand a similar set of questions for half-BPS geometries in $AdS_5\times S^5$. These geometries, constructed by Lin, Lunin and Maldacena \cite{Lin:2004nb}, which we will call LLM geometries, have similar properties to the fuzzball microstates. They are horizon free geometries, they have bubbles and fluxes and come in large families. These solutions can have different topologies. Part of their simplicity is that their dual states in the boundary CFT are easy to describe. The geometries are dual to a very simple sector of ${\cal N}=4$ SYM that can be described in terms of 2D incompressible droplets of free fermions \cite{Berenstein:2004kk}.  The droplet picture also arises directly from the supergravity solutions, so one can actually have a two-dimensional picture of the spacetime that encodes all the data necessary to recover the ten-dimensional geometry \cite{Lin:2004nb}. We will call this two-dimensional surface the LLM plane.

These solutions do not admit black holes and have no Wick rotated Euclidean geometry related to them. They can still be coarse-grained. When one does so, the droplets become grey: they can be a statistical ensemble of filled or unfilled regions of the LLM  plane, so the local droplet density on the LLM plane instead of being zero or one can be any number in between. The corresponding greyscale supergravity solutions have naked singularities. Essentially, there are not enough half-BPS states to give an entropy of order $N^2$ and generate a horizon with an area of the right size in geometric units.
One can think of these as incipient black holes \cite{Balasubramanian:2018yjq} and put a stretched horizon around the singular locus. It should also be noted that in these geometries the topology of the spacetime is not a quantum observable. The topology of LLM geometries can be changed by the superposition of a very large number of states \cite{Berenstein:2017abm}. This fact guarantees that the topology is not a quantum observable \cite{Berenstein:2016pcx}. The notion of topology depends on the coarse-graining, but for large enough bubbles this is not a major issue. This is similar to arguing that the topology of the Euclidean spacetime is not characterized by a {\em conserved} quantum number, as we can take extreme limits where the small black holes evaporate and change topology. 

Armed with explicit solutions of the LLM geometries, we aim to ask to what extent an individual LLM microstate might behave as a black hole. The whole point is to argue that being a black hole in the dual theory is more a matter of degree, rather than an expression of the quantum numbers of the state.
Therefore, even in these cases where one barely has a notion of an incipient black hole, where even the notion of the horizon is not available, it is interesting to ask if light gets trapped. We also want to come to a simplified picture of what is responsible for the {\em trapping mechanism} and ask about the existence of tidal forces that might pull objects apart \cite{Tyukov:2017uig,Bena:2018mpb,Bena:2020iyw,Martinec:2020cml} (see also \cite{Eperon:2016cdd,Marolf:2016nwu,Eperon:2017bwq,Bena:2020yii}).

To test if light is trapped, we consider solving for the null geodesics of the LLM geometries.
A particularly useful observation we have in this paper is that there are null geodesics whose motion is restricted to lie in the LLM plane. The motion is that of a particle in two dimensions in the presence of non-trivial magnetic fields. These are restricted to be either in the filled or empty region of the LLM plane.

The picture that appears is very appealing.  For geodesics in the empty region,  the droplets are obstacles in the path of the null geodesics that they need to avoid. Essentially, the geodesics are reflected from the droplets, similar to a problem of chaotic billiards. 
The fact that we see chaos in the dynamics can be used as an analogue of strong tidal forces: nearby null geodesics will separate exponentially fast from each other. A notion of the trapping mechanism can be extrapolated to the case where there are many droplets. The null particles in that case 
are better thought of as moving through a dense medium with multiple scattering centers. A typical microstate can be argued to have many droplets of random shapes.
If the results of collisions are fairly random, the process of getting outside a region is analogous to diffusion. This mechanism can explain simply why something that falls in might take a very long time to get out. Some such geodesics might be completely encircled so they can never get out. Such modes could in principle become internal excitations of the putative incipient black hole that are very hard to tease out by an outside observer.

The paper is organized as follows. In section \ref{sec:LLMgeo} we briefly summarize the ten-dimensional LLM geometries to establish notation. In section \ref{sec:Dynamics}, we describe the dynamical system of our interest.
We turn the problem of studying null geodesics into a Hamiltonian dynamics problem in very few variables. The main reason to do so is that the LLM geometries have multiple symmetries that lead to conserved quantities. In the Hamiltonian formulation, these constants of motion are angular momenta and can be frozen to reduce the dimensionality of the problem. We will also end up studying a subset of the geodesics that lie in the LLM plane. In section \ref{sec:OneDisk}, we describe the geodesic motion in vacuum $AdS_5\times S^5$, where the LLM data is one round disk case as a warm-up. Part of the purpose is to later compare with a geometry with more droplets. 
In section \ref{sec:Numerics} we present numerical results that characterize the chaotic behavior of the three disks configuration. Because the work we do is numerical, we study a configuration with three disks as an example, but the main results are expected to be generic.

\section{LLM geometries}
\label{sec:LLMgeo}
In this section, we briefly review the bubbling $AdS$ geometries constructed in \cite{Lin:2004nb} to establish notation. 

Motivated by the AdS/CFT correspondence, Lin, Lunin and Maldacena (LLM) explored asymptotically $AdS_5\times S^5$ 
half-BPS IIB supergravity solutions. These solutions  preserve an SO$(4)\times$SO$(4)\times\mathbf{R}$ symmetry.
This residual symmetry of configurations is easiest to understand in the dual CFT formulation \cite{Berenstein:2004kk}, where only one of the complex scalars of ${\cal N}=4$ is excited. The field theory is analyzed in the $S^3\times {\mathbb R}$ cylinder, whose ground state is dual to $AdS_5\times S^5$ in global coordinates.
The unbroken R-symmetry is $SO(4)$ and the configuration is also spherically symmetric on the spatial $S^3$ of the boundary. The BPS constraint adds one extra conserved charge, identified with $\mathbf{R}$. 

The upshot of the field theory analysis is that the half-BPS configurations of the SYM theory are described in the weak coupling limit in terms of eigenvalues of a complex matrix (the Gauss' law constraint turns the matrix variable describing the half-BPS states into a normal matrix). The real and imaginary parts of the eigenvalues are conjugate variables to each other. The measure terms from gauge fixing make these eigenvalues behave like fermions.
The effective system describing the dynamics of this reduced set of states is that of droplets of fermions in the presence of a magnetic field, when reduced to the lowest Landau level. Correlators in the half-BPS sector can be computed exactly for these states by combinatorial methods \cite{Corley:2001zk}. 
More recently, other tools have been used to simplify how these results can be obtained \cite{Berenstein:2022srd} by using coherent state methods and integral representations that can be evaluated by localization.

With these symmetry considerations Lin, Lunin and Maldacena found the following general solution, in type IIB supergravity, for the metric,
\begin{eqnarray}
\label{eq:metricLLMcartesian}
	h^2 ds^2=-(dt+V_adx^a)^2+h^4(d\xi^2+dx_a dx^a)+\left(\frac{1}{2}-z\right)d\tilde{\Omega}_3^2+\left(\frac{1}{2}+z\right)d\Omega_3^2.
\end{eqnarray}
The $SO(4)\times SO(4) \times \mathbf{R}$ symmetry is manifest, with all functions being independent of $t$ and the $SO(4)$ symmetries realized by isometries of the two 3-spheres $\tilde{\Omega}_3$ and $\Omega_3$. 
The functions $z$ and $h$ depend only on the three variables $\xi, x^{1,2}$.

The $h$ function is related to $z$ by,
\begin{eqnarray}
\label{eq:h-and-G}
h^2=\frac{1}{\xi}\sqrt{\frac{1}{4}-z^2},
\end{eqnarray}
while the $V_a$ functions obey the following equations in terms of $z$ and $\xi$,
\begin{eqnarray}
\label{eq:zequations}
	\partial_\xi V_a=\frac{\epsilon_{ab}\;\partial_bz}{\xi},\qquad \partial_aV_b=\frac{\epsilon_{ab}\;\partial_\xi z}{\xi}+\partial_bV_a.
\end{eqnarray}
This means that the solution is completely characterized by the single function $z=z(\xi,x, y)$, where $x^a=\{x,y\}$ and the range of the coordinates is $\xi\in(0,\infty)$ and $x^a\in(-\infty,\infty)$. That is,  the domain of the $z$ function is an upper half space in three dimensions.

Besides the metric, the background is supported by a five-form in the RR sector, which together with the metric \eqref{eq:metricLLMcartesian} is a solution of type IIB supergravity if the following linear PDE for $z$ is satisfied,
\begin{equation}\label{Generating function}
    \partial_a\partial_a z+\xi\partial_\xi \left(\frac{\partial_\xi z}{\xi}\right)=0.
\end{equation}

The solutions are tuned with precise boundary conditions, $z\to\pm 1/2$ as $\xi\to0$, in such a way that the metric $ds^2$ does not contain conical singularities. One gets a  'Black' and 'White' colouring $xy$-plane for each geometry defined by,
\begin{equation}
\label{eq:Boundary conditions}
   \text{White:}\quad  \lim_{\xi\to 0} z(\xi, x, y)=\frac{1}{2}, \qquad \text{Black:}\quad  \lim_{\xi\to 0} z(\xi, x, y)=-\frac{1}{2}.
\end{equation}
The convention of colors can also be thought of as  `White' being a region with no fermions and  `Black' a filled region with the maximum density of fermions that is allowed. This $\xi=0$ surface will be called the LLM plane and is parametrized by $x^{1,2}$.
The smoothness restriction arises because we need $h$ to be finite in the LLM plane. More precisely, it is  $\lim_{\xi\to 0} h^{-1} \neq0$ that must be non-vanishing. Otherwise, both spheres shrink to zero size at $\xi=0$. The boundary between the two saturated color regions is allowed as there the metric becomes that of a non-singular plane wave \cite{Lin:2004nb}.
The two coloring of the LLM plane is that of an incompressible fluid and reproduces exactly the ${\cal N}=4$ free fermion picture.

Notice that the boundary conditions of $z$ on the $\xi=0$ plane play an important role as it describes the size of the two 3-spheres. In the $z\to-1/2$ regions the 3-sphere $\Omega_3$ shrinks to zero as $\xi\to 0$, whilst in the $z\to+1/2$ regions the 3-sphere $\tilde{\Omega}_3$ shrinks to zero as $\xi\to 0$. These are easy to see in \eqref{eq:metricLLMcartesian} if $h$ is finite. The radii of the spheres have additional factors of $ 1/2\pm z$ in the metric.

Note that these boundary conditions ensure smoothness of the solution \eqref{eq:metricLLMcartesian} in the entire space. If one accepts greyscale solutions, the limit $\xi \to 0$ has singularities. The function $z$ is then in the range $(-1/2,1/2)$. These are the only values that are compatible with causality \cite{Caldarelli:2004mz} (other values lead to closed timelike curves). 
The effective reason for these bounds in the dual ${\cal N}=4$ SYM is that the fermions must satisfy the Fermi exclusion principle. The density of fermions has an upper bound, and similarly, cannot be negative.

\subsection{The one disk and multi disks geometries}

The simplest configuration is the geometry described by the $AdS_5\times S^5$ spacetime itself. 
In the LLM plane, this is described by a single disk. This solution is described by the following $z$ function,
 \begin{equation}
\begin{split}
\label{eq:zOnedisc}
	z=&\frac{r^2+\xi^2-R^2}{2\sqrt{(r^2+\xi^2+R^2)^2-4r^2R^2}},\qquad\text{with}\qquad z(\xi\to0)=\left\{ \begin{array}{cccrcl}
                       -\frac{1}{2} & r<R, &\\
                                 \frac{1}{2}    &~~ r>R,&                                             \end{array}
\right.
\end{split} 
\end{equation}
where $r^2={x}^2+{y}^2$ and $R$ is the radius of the disk. When we consider more general disk configurations at the boundary $\xi\to0$,
 the rotational symmetry in the $x_a$ plane is lost.

We can easily find generalizations to geometries with n-disks. We do this by summing solutions for 
various disks. This is where the fact that we need to solve a linear PDE does its work: we just superpose (sum) solutions and fix a constant so that the empty region is at the right value of $z$.
We write these geometries in  Cartesian coordinates and the $z_T$ function is obtained as the sum of each individual $z_i$ function, 
\begin{equation}
\begin{split}
\label{eq:zdiscI}
	z_i&=\frac{r_i^2+\xi^2-R_i^2}{2\sqrt{(r_i^2+\xi^2+R_i^2)^2-4r_i^2R_i^2}},\\
\text{with}\ \ & z_T=\sum_i^n\left(z_i+\frac{1-n}{2n}\right)\ \
	\text{and}\ z_T(\xi\to0)=\left\{ \begin{array}{cccrcl}
                       -\frac{1}{2} & r<R, &\\
                                 \frac{1}{2}    &~~ r>R,&         \end{array}
\right.
\end{split} 
\end{equation}
where $n$ is the number of disks and the subscript $i$ labels the $i$th-disk. The $i$th-disk has its center at coordinates at $\{x_{i},y_{i}\}$ and radius $R_i$, and we define $r_i^2=(x-x_{i})^2+(y-y_{i})^2$. The constant factor $\frac{1-n}{2n}$ guarantees that the boundary conditions $z=\pm 1/2$ are satisfied. 

The $V_a$ functions that solve the equations \eqref{eq:zequations} with the $z_T$ function given by \eqref{eq:zdiscI} are, 
\begin{eqnarray}
\label{eq:vs}
V_{a}&=&\frac{\epsilon_{ba}}{2}\sum_{i}^n\frac{(r_i^2+\xi^2+R_i^2)}{\sqrt{(r_i^2+\xi^2+R_i^2)^2-4r_i^2R_i^2}}\partial_{x_b}(\log{r_i}).%\\
% V_{x}&=&-\frac{1}{2}\sum_{i}^n\frac{(r_i^2+\xi^2+R_i^2)\;\partial_{y}(\log{r_i})}{\sqrt{(r_i^2+\xi^2+R_i^2)^2-4r_i^2R_i^2}},\qquad V_{y}=\frac{1}{2}\sum_{i}^n\frac{(r_i^2+\xi^2+R_i^2)\;\partial_{x}(\log{r_i})}{\sqrt{(r_i^2+\xi^2+R_i^2)^2-4r_i^2R_i^2}}.\nonumber\\
\end{eqnarray}
This way, we have a readily available explicit list of interesting geometries to consider. 

\section{General dynamics of the system}
\label{sec:Dynamics}

In order to study the dynamics of the system we compute the Hamiltonian from the metric \eqref{eq:metricLLMcartesian}. We call the null geodesic affine time parameter of the dynamical problem $\tau$.
We start from the Lagrangian
\begin{equation}
L = \frac 12  g_{\mu\nu} \dot x^\mu \dot x^\nu 
\end{equation}
and compute the Legendre transform.
Null geodesics are interpreted as the geometric limit of massless particles. In the type IIB string theory, these massless particles are all part of the gravity multiplet in 10 
dimensions, so they are captured by all the possible supergravity modes. Any statement about null geodesics has consequences for wave solutions in supergravity.

 The spherical symmetry in $\Omega_3$ and $\tilde \Omega_3$ automatically lets us introduce conserved angular momenta $L,\tilde L$. Also, the time translation invariance in the $t$ variable lets us introduce $E$ as the conserved conjugate momenta of $t$. The null geodesic constraint will have the Hamiltonian function vanishing (${\cal H}=0$). 
 The Hamiltonian (rescaled by $h^2$) for the dynamics of the remaining three variables is the following
\begin{equation}
	\begin{split}
		2h^2\;\mathcal{H}
		=&P_\xi^2+(P_{x}+EV_{x})^2+(P_{y}+EV_{y})^2+h^4\left(\frac{2L^2}{\left(1+2z\right)}+\frac{2\tilde{L}^2}{\left(1-2z\right)}-E^2\right),
\end{split}
\label{eq:hamiltonianCart}
\end{equation}
where the two coordinates $x^{a}$ have been renamed to $x,y$ and the third coordinate $\xi$ of the LLM metric has remained the same. The $P_{x,y}$ are conjugate to $x,y$ and $P_\xi$ is conjugate to $\xi$.

We want to study geodesics that reach the incipient black hole region $\xi\simeq 0$.
One can check that the geodesics that reach the LLM plane $\xi\to0$ must have an angular momentum vanishing in one of the two possible spheres. Which angular momentum vanishes depends on which sphere goes to zero size at the intersection. This mathematical fact generates an angular momentum repulsion barrier. Also notice that the $\xi=0$ region is a locus of unbroken rotation symmetry for one of the two spheres, depending on which sphere degenerates. This indicates that it is possible to find solutions to the equations of motion that stay completely in this locus. These will have $P_\xi=0$ and $\xi=0$ and one of the angular momenta vanishing. Our main computational effort will be to study this subset of solutions to the null geodesics that are strictly confined to the LLM plane.

 In such limit, the system is reduced to a four-dimensional phase space with the following Hamiltonian,
\begin{equation}
	\begin{split}
	\mathcal{H}_0
		=&\left.\frac{1}{2h^2}\left((P_{x}+EV_{x})^2+(P_{y}+EV_{y})^2+h^4\left(L^2-E^2\right)\right)\right|_{\xi\to0}.
\end{split}
\label{eq:hamiltoniany0}
\end{equation}
where both $L,E$ are constants of motion. Therefore, the interesting dynamics takes place only in the $x,y$ coordinates.

Null geodesics then satisfy the following constraint,
\begin{equation}
	\begin{split}
	\left.(P_{x}+EV_{x})^2+(P_{y}+EV_{y})^2+h^4\left(L^2-E^2\right)\right|_{\xi\to0}=0,
\end{split}
\label{eq:nullconstraint}
\end{equation}
which can be expressed in terms of the velocities in a quite simple way as,
\begin{equation}\label{CircularConstraint}
    E^2-L^2=\dot{x}^2+\dot{y}^2.
\end{equation}
This form of the constraint suggests that the motion should be very simple.
The constraint tells us that $E>L$ (this is a classical limit of the BPS constraint). When $E=L$, there is no motion in the $x,y$ coordinates. There is one of these stationary solutions for any empty point in the LLM plane \footnote{These are also related to the zero size limit of BPS open strings discussed in \cite{Berenstein:2020jen}.}.

The equations of motion derived from the Hamiltonian \eqref{eq:hamiltoniany0} that satisfy the requirement \eqref{eq:nullconstraint} are,
\begin{equation}
    \begin{split}
	\label{eq:eomCar}
\dot{x_a}&=\left.\frac{P_{x_a}+E\;V_{a}}{h^2}\right|_{\xi\to0},\\
\dot{P}_{x_a}&=\left.(E^2-L^2)\partial_{x_a}h^2-\frac{E}{h^2}\left((P_{x}+EV_{x})\partial_{x_a}V_{x}+(P_{y}+EV_{y})\partial_{x_a}V_{y}\right)\right|_{\xi\to0},
 \end{split}
\end{equation}
for $x_a=(x,y)$. The $h^2$ and $V_{a}$ functions in the $\xi\to0$ limit are given by the following explicit expressions
\begin{equation}
     \begin{split}
    \left.h^2\right|_{\xi\to0}&=\sqrt{\sum^n_{i=1} \left(\frac{R}{|r_i^2-R^2|}\right)^2},\qquad
    \left.V_{a}\right|_{\xi\to0}=\frac{1}{2}\epsilon_{ba} \sum_{i=1}^n\frac{(r_i^2+R^2)}{|r_i^2-R^2|}\partial_{x_b}(\log{r_i}).
     \end{split}
\end{equation}
One can see that the explicit equations of motion in this limit are somewhat unwieldy, suggesting that finding analytical solutions to this system of differential equations becomes a complicated task. The quantities $E V_x, E V_y$ in equation \eqref{eq:hamiltoniany0} act as an effective magnetic field potential on the LLM plane.
The dynamical system is that of particles subject to a non trivial metric, a non-trivial potential proportional to $L^2-E^2$ and a non-trivial magnetic field that is time independent. 

This low dimensionality of the problem makes it easier in the end to visualize the solutions.  
In the next section we will show a numerical analysis of our system.

\section{The global $AdS_5\times S^5$ geometry}\label{sec:OneDisk}
In order to contrast with more elaborate configurations, in this section we present the dynamics of the simplest case in the LLM geometries: the $AdS_5\times S^5$ vacuum itself. 
The main idea is to understand the null geodesics of interest to us in the LLM coordinates, rather than the more conventional global $AdS$ coordinates. The geometry here is determined by one disk with radius $R$. The disk on the LLM plane has rotational symmetry, which is associated with the conserved angular momentum $P_\theta$. There is also the conserved energy of the particle $E$ in these coordinates. In the plane wave limit geometry \cite{Berenstein:2002jq}, the quantity $E$ is identified with $E=\Delta -J$, where $\Delta$ is the global energy (the one that measures anomalous dimensions of operators) and $J$ is the R-charge carried by the LLM geometry.  This is a true statement in general.
The LLM geometries satisfy the constraint $E_{back}=\Delta -J=0$. This is the BPS constraint. The geometries have vanishing energy with respect to the LLM time variable. This is also part of the dual field theory setup \cite{Berenstein:2004kk}.
The $P_\theta$ momentum is the additional amount of R-charge $J$ carried by the null particle whose geodesic motion we are analyzing. 

The general Hamiltonian (\ref{eq:hamiltoniany0}) mapped to  polar coordinates for this particular case is
\begin{equation}
	\begin{split}
	\mathcal{H}_0
		=&\left.\frac{1}{2h^2}\left(P_{r}^2+\frac{(P_{\theta}+EV_{\theta})^2}{r^2}+h^4\left(L^2-E^2\right)\right)\right|_{\xi\to0}.
\end{split}
\label{eq:hamiltonian}
\end{equation}
The canonical momenta are given by
\begin{equation}
    P_r=h^2 \dot{r}, \quad h^2 E=\dot{t}+V_{\theta}\dot{\theta}, \quad P_{\theta}=-EV_{\theta}+h^2r^2\dot{\theta}.
\end{equation}
The additional rotational symmetry on the LLM plane transforms this problem into a one-dimensional system, as $P_\theta$ is conserved. Furthermore, since the Hamiltonian is time-independent, then the system is integrable. This result is also true for generic concentric circle geometries. The integrability is for null geodesics inside the LLM plane, which is the reduction we are studying in this paper. 

In turn, the null constraint \eqref{eq:nullconstraint}, which is also ${\cal H}_0=0$, can be recast as a condition over the domain of  $r$, as $P_r$ must be real.
We can solve for $P_r$ as follows:   
\begin{equation}
    P_{r}=\pm\sqrt{h^4\left(E^2-L^2\right)-\frac{(P_{\theta}+EV_{\theta})^2}{r^2}}.
\end{equation}
All the terms on the right hand side depend on $r$ and the constants of motion. 
Notice that, for certain values of the conserved momenta $(L,E,P_\theta)$, the quantity $P_r$ vanishes, giving us turning points for $r$. 

In the $\xi\to0$ limit, one can also determine the explicit form of functions $h^2$ and $V_{\theta}$ in polar coordinates,  
\begin{equation}
     \left. h^2\right|_{\xi\to0}=\frac{R}{|r^2-R^2|}, \quad  \left. V_{\theta}\right|_{\xi\to0}=\frac{1}{2}\left(\frac{r^2+R^2}{|r^2-R^2|}-1\right).
\end{equation}

\begin{figure}[ht]
    \centering
    \includegraphics[width=10cm]{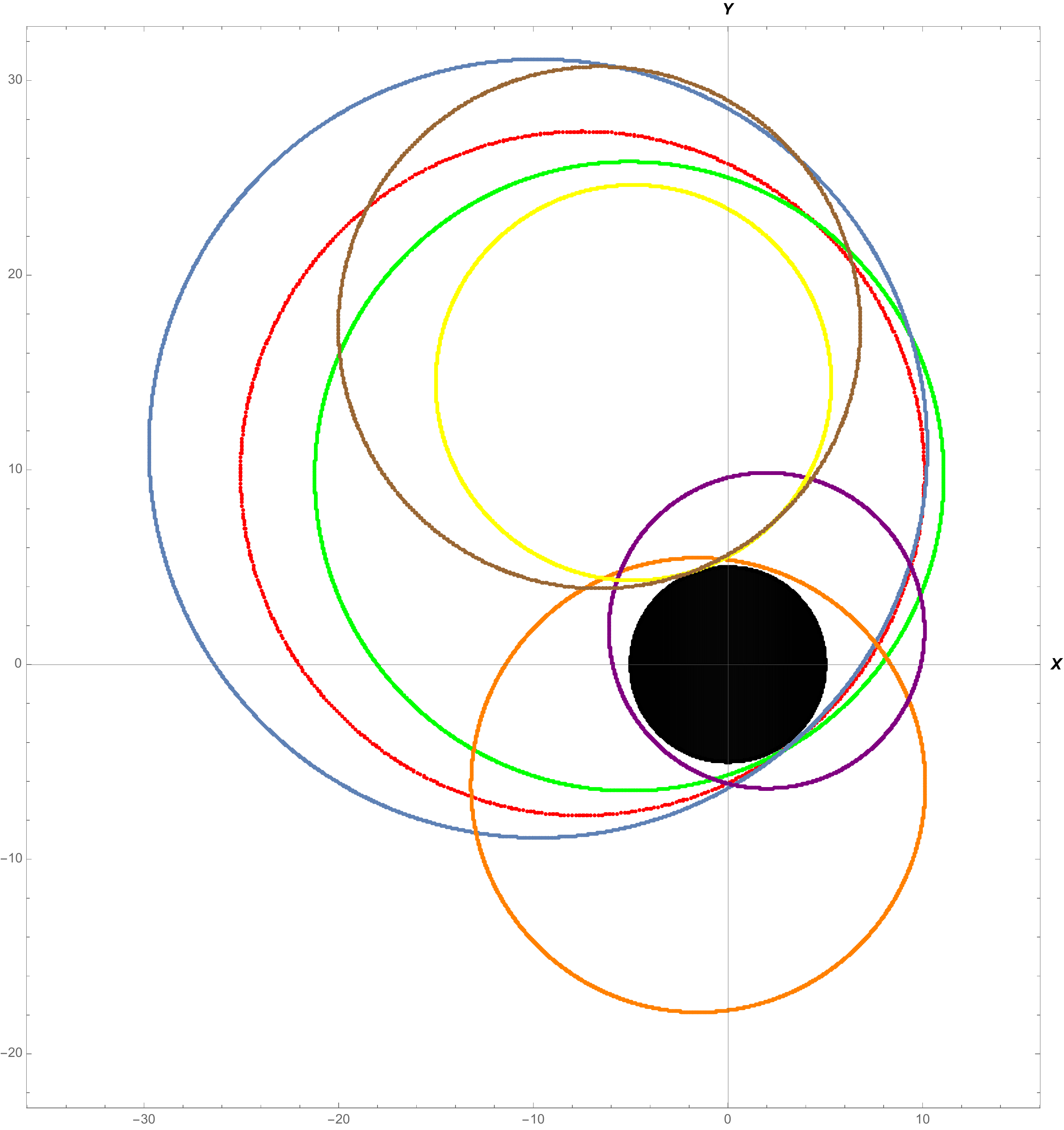}
    \caption{ The disk of $R=5$ centered at the origin is represented by a black circular region. The rest of the circles with different colors represent geodesics with different initial conditions. Each orbit has a circle-like shape, but are not exactly circular orbits. Some of the geodesics (brown and yellow) do not orbit the central black disk. }
    \label{fig:OneDisk}
\end{figure}

In Figure \ref{fig:OneDisk} we show a set of orbits where we consider a disk centered at the origin with radius $R=5$. We consider geodesics with $E=20$ and $L=5$, but with different values for initial positions and momenta.

Notice also that there exist trivial analytic solutions in the LLM plane with $E=L$ (the general BPS solutions we argued in the previous section). These solutions have a nontrivial angular momentum $P_\theta=-E V_\theta$ and $P_r=0$.
In these coordinates, these null geodesics are all stationary, but they still move in the time direction $t$ on the one 3-sphere on which they have angular momentum. 

\section{Numerical results and graphs for one example: three disks}
\label{sec:Numerics}

Now we are ready to study a more general case. In general, black holes are expected to lead to chaotic behavior in the dynamics \cite{Shenker:2013pqa,Maldacena:2015waa}. One would expect that for an LLM geometry to behave similarly to a black hole, a prerequisite is to have the geodesic solutions be chaotic. Given the high degree of symmetry of the solutions, this is not guaranteed in general. 
The results of \cite{Chervonyi:2013eja} suggest that most LLM geometry solutions will lead to a non-integrable system, even for null geodesics. These arguments are based on the separability of the associated Hamilton-Jacobi problem. Here we verify explicitly that there is chaos. 
This fact by itself does not mean that a system behaves as a black hole, but it serves as a prerequisite for that statement to be true. For example, in the $AdS_5\times T^{1,1}$ geometry, there is chaos for string motions \cite{Basu:2011di}, but the geodesic problem itself is integrable.
Clearly, there is no black hole. In this section, we want to verify explicitly the claim of the existence of chaos in one simple configuration with multiple disks. 

Because we expect chaos,  we will verify the presence of chaos numerically. We choose one example of a geometry with enough features to be interesting, but with the expectation that some of the features of motion are generic.

We consider one example of a configuration with three disks of the same radius $R=5$, whose centers are located at the following positions in the LLM plane
\begin{equation}\label{PositionsDisks}
    (x_{1},y_{1})=(0,0),\qquad (x_{2},y_{2})=(20,0),\qquad (x_{3},y_{3})=(10,20).
\end{equation}
Our initial setup considers geodesics of energy $E=20$ and angular momentum $L=5$. These are chosen to be of order one.
The only parameter that physically matters is $ L/E$, as solutions of null geodesics have a rescaling property: the Hamiltonian is a purely quadratic function of momenta. 
The null geodesics with momenta all rescaled by the same factor will follow the same trajectory in the position coordinates, but they will evolve faster or slower in the affine parameter $\tau$.

\begin{figure}[ht]
    \centering
    \includegraphics[width=8cm]{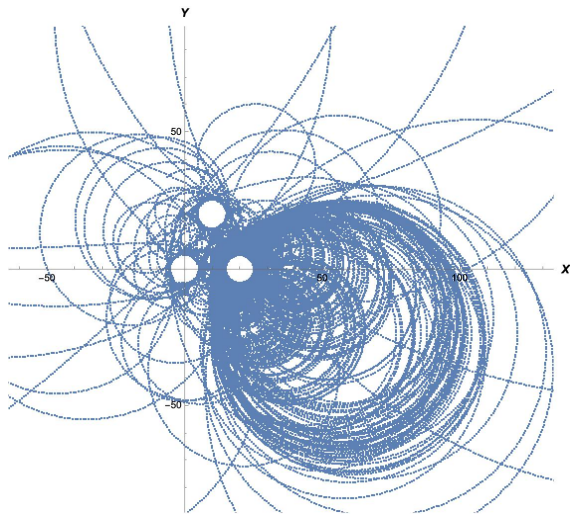}
    \caption{ The orbit followed by a single null geodesic in the three disk geometry projected to the LLM plane.}
    \label{fig:ThreeDisksOrbit}
\end{figure}

Figure \ref{fig:ThreeDisksOrbit} shows an orbit plot for a given generic initial condition in position and momenta. We see the location of the three disks demarcated in the plot by the region that the null geodesics are avoiding. Comparison with Figure \ref{fig:OneDisk} shows that this new orbit is significantly more complicated than in the one disk case. The rotational symmetry on the $xy$-plane from before is now lost due to the additional disks. We see that the motion is highly irregular: the orbit spends most of its time circling around a single disk, but jumps occasionally to motion around the other disks. Less frequently, the orbit travels far outside of all of the disks, as seen in the figure.

The orbits are prevented from reaching the inside of the disks by the angular momentum $L$. As we explained earlier, the two 3-spheres degenerate in the LLM plane and there is an angular momentum repulsion from the filled disk region. In that sense, we can think of the disks as impenetrable obstacles that the trajectory must be reflected from. This is similar to a billiards problem. The main difference is that there is also an effective magnetic field present. This effective magnetic field arises because the metric is not static in the $t$ coordinate.

 The trajectory appears to bounce off these disks. Some parts of the trajectory also travel far out in the radial direction, closer to the boundary of $AdS$. 
Since the trajectory only has angular momentum in the $SO(4)$ rotational group of the $S^3$ on the boundary, these can in principle reach infinity if they stop carrying angular momentum with respect to the $U(1)$ charge of rotations in the LLM plane. That is, if $P_\theta$ becomes very small (the angular momentum $P_\theta$ is not conserved, but should become an
adiabatic invariant in trajectories of the region very far from the disks). The rotational breaking of symmetry makes it in principle possible to do this (we can have jumps in $P_\theta$
between different portions of a geodesic).

\begin{figure}[ht]
     \centering
     \begin{subfigure}[b]{0.65\textwidth}
         \centering
         \includegraphics[width=\textwidth]{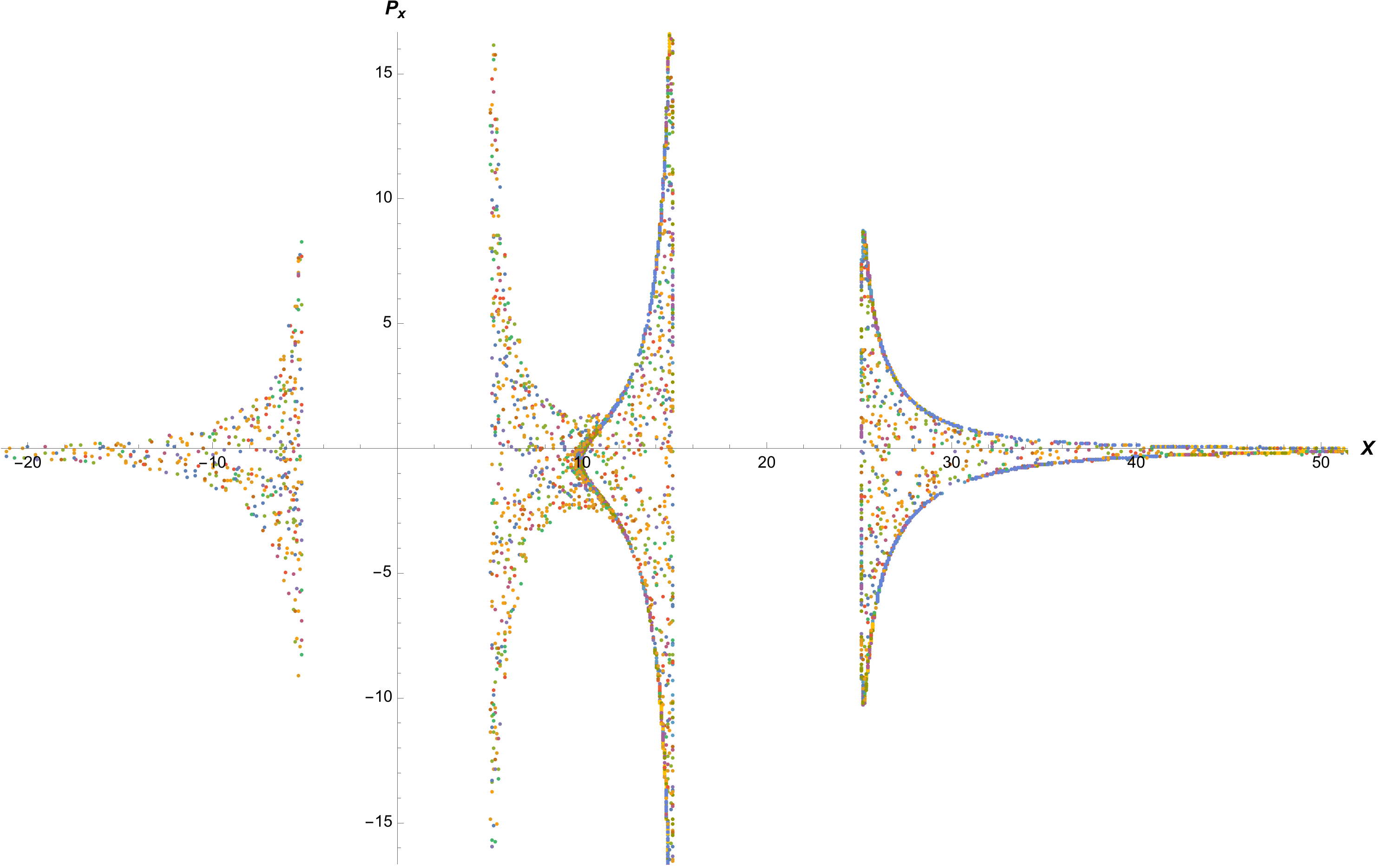}
         \caption{Poincar\'e section for $y=0$.}
         \label{fig:PoincareY0}
     \end{subfigure}
     \hfill
     \begin{subfigure}[b]{0.65\textwidth}
         \centering
         \includegraphics[width=\textwidth]{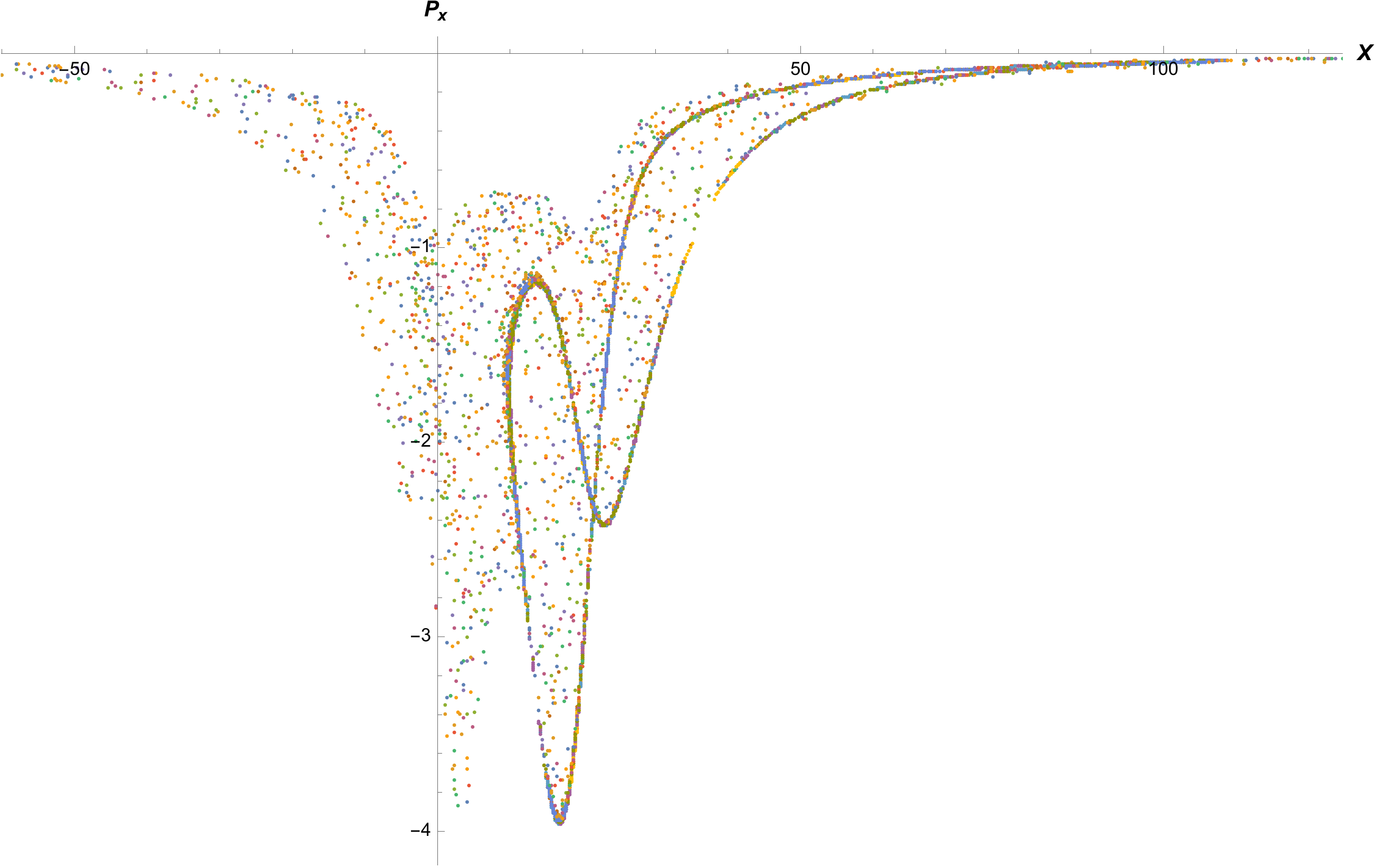}
         \caption{Poincar\'e section for $y=-8$.}
         \label{fig:PoincareYm4}
     \end{subfigure}
        \caption{Poincar\'e sections, $x$ vs $P_x$, for some values of $y$. The exterior white regions represent the region disallowed by the null constraint, where the position $x$ is roughly inversely proportional to the momentum $P_x$. The points fill out completely the allowed region in the phase space as we expect for chaotic motion. 
        In the first plot, the two vertical white strips represent the locations of the two disks at $y=0$, which are forbidden. In the second plot, the two vertical white lines are gone since the cut is moved to $y=-8$.}
        \label{fig:PoincareSectionYfixed}
\end{figure}

Since the rotational symmetry on the plane is lost, the system that was originally integrable for one disk becomes chaotic for three disks. Due to the low effective dimension of the problem (a two-dimensional phase space), we can visualize chaos with the aid of Poincar\'e sections.
 In Figure \ref{fig:PoincareSectionYfixed}, the Poincar\'e sections with the $y$ variable fixed are shown. They clearly display chaos.

%\begin{figure}[ht]
%     \centering
%     \begin{subfigure}[b]{0.47\textwidth}
%         \centering
%         \includegraphics[width=\textwidth]{PoincareSection x=10.pdf}
%         \caption{Poincare section for $x=10$}
%         \label{fig:PoincareX10}
%     \end{subfigure}
%     \hfill
%     \begin{subfigure}[b]{0.47\textwidth}
%         \centering
%         \includegraphics[width=\textwidth]{PoincareSection x=20.pdf}
%         \caption{Poincare section for $x=20$}
%         \label{fig:PoincareX20}
%     \end{subfigure}
%        \caption{Poincare sections, $y$ vs $P_y$ for some fixed values of $x$. In the first plot, the cut is at $x=10$ so that the cut does not hit a disk. In the second plot, the cut is at $x=20$ where the white line represents the location of one of the disks.}
%        \label{fig:PoincareSectionXfixed}
%\end{figure}

The main characteristic of chaotic dynamics is the butterfly effect \cite{Shenker:2013pqa}: exponential sensitivity to initial conditions. This is checked by demonstrating that the system has a non-trivial Lyapunov spectrum (Lyapunov exponents that don't vanish). 
To compute Lyapunov exponents, we choose the following initial conditions $(x^{(0)},y^{(0)})=(10,3)$ and $(P_{x}^{(0)},P_{y}^{(0)})=(\frac{152785}{856086},\frac{25\sqrt{1695}}{616})$, so that we can have arbitrary precision if necessary. Since the three disks configuration has a four-dimensional phase space, we report the four Lyapunov exponents associated with the configuration, checking that they are numerically in accordance with the following pattern numerically $\lambda_1> \lambda_2>-\lambda_2>-\lambda_1$, and that $\lambda_2\simeq 0$ due to there being a constant of motion. The pattern arises because the chaotic motions respect the symplectic structure of phase space. Computing the exponents numerically, we find that
\begin{equation}
     \{\lambda_i\}=\{0.4484, 0.02104, -0.02109, -0.4483\}.
\end{equation}

\begin{figure}[ht]
\centering
\includegraphics[width=13cm]{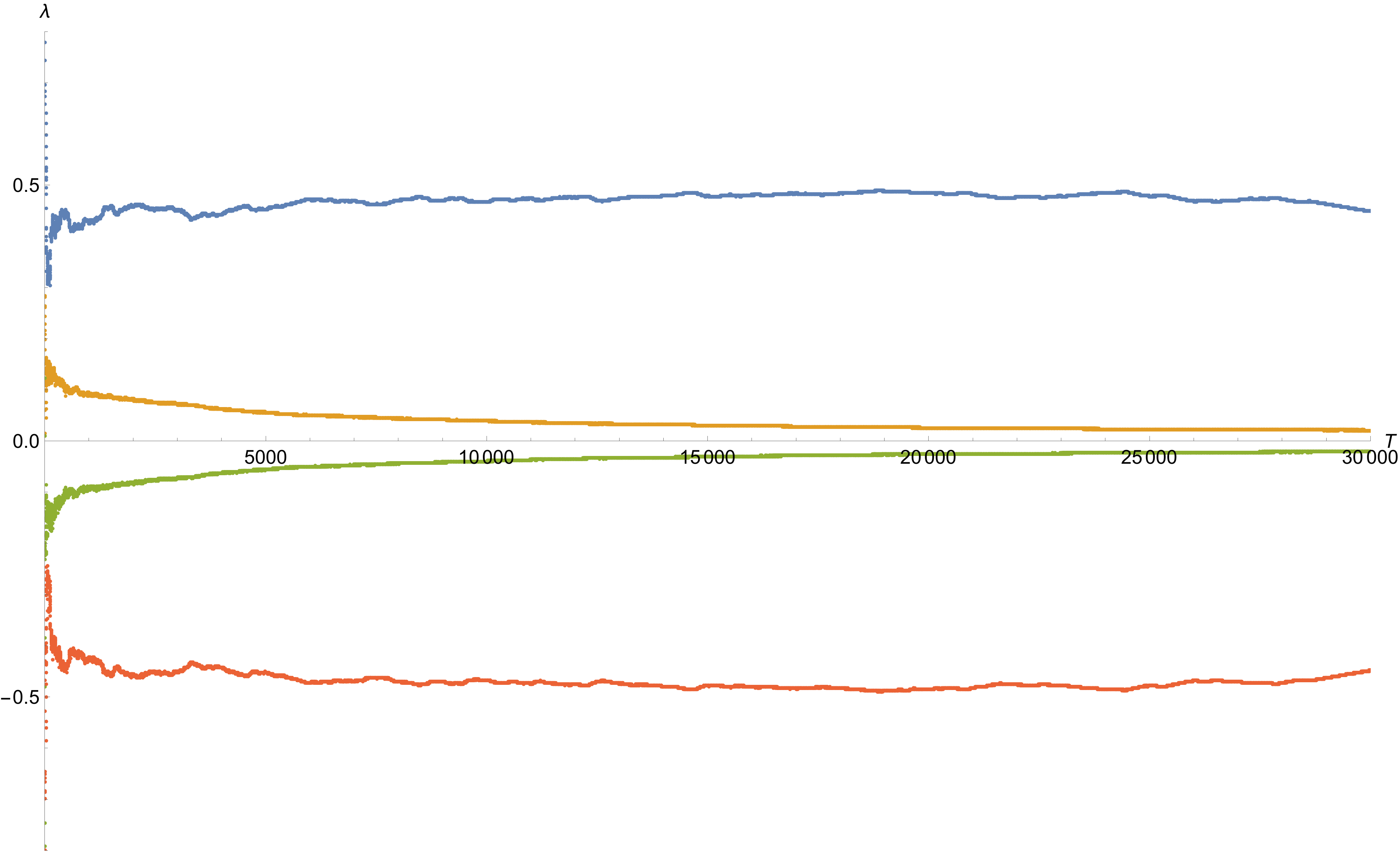}
\caption{The maximum Lyapunov exponent $\lambda_{max}=0.4484$ is represented by the blue line.}
\label{fig:LyapunovRepTrajectory}
\end{figure}

The maximum Lyapunov exponent is positive and  $\lambda_{max}\sim0.4484$ is represented by the blue data in Figure \ref{fig:LyapunovRepTrajectory}. 
We can use the difference from zero of the Lyapunov exponents as an error bar. This method would indicate that one expects the largest exponent to have an error of order  $5\%$.  
We expect two of the Lyapunov exponents to vanish. This is supported by our analysis with two of the exponents slowly converging toward zero. The slow convergence comes from the form of the differential equations we are integrating. The equations become stiff near disk boundaries, setting limits as to how far we can integrate up to.

\subsection{Geodesics trapped in the interior region formed by the disks}\label{sec:trapped}

The goal of our analysis is to observe trajectories that are almost trapped and take a very long time to escape to infinity. One way to create this behavior is to make the exits of some regions very narrow by placing our disks close to each other. Due to the angular momentum repulsion effect, this causes a trajectory to bounce back and get trapped for an extended period.
For our setup, we choose the same center locations for three disks as before (see (\ref{PositionsDisks})), but we increase the radius of the disks such that they are almost touching each other. This causes a geodesic starting in the interior region formed by the three disks to get effectively trapped.  
We choose $R=\frac{9999}{1000}$ for each disk. The energy is kept the same at $E=20$, but the angular momentum is changed to $L=19$. This is close to the BPS limit. 
The initial condition of the representative trajectory is $(x_0,y_0)=(10,3)$ and $v_{x_0}=1$, $v_{y_0}>0$. We only show some points for a very long time trajectory. If we filled the points of the trajectory, the region where the null geodesic is trapped would seem filled.

\begin{figure}[ht]
    \centering
    \includegraphics[width=12cm]{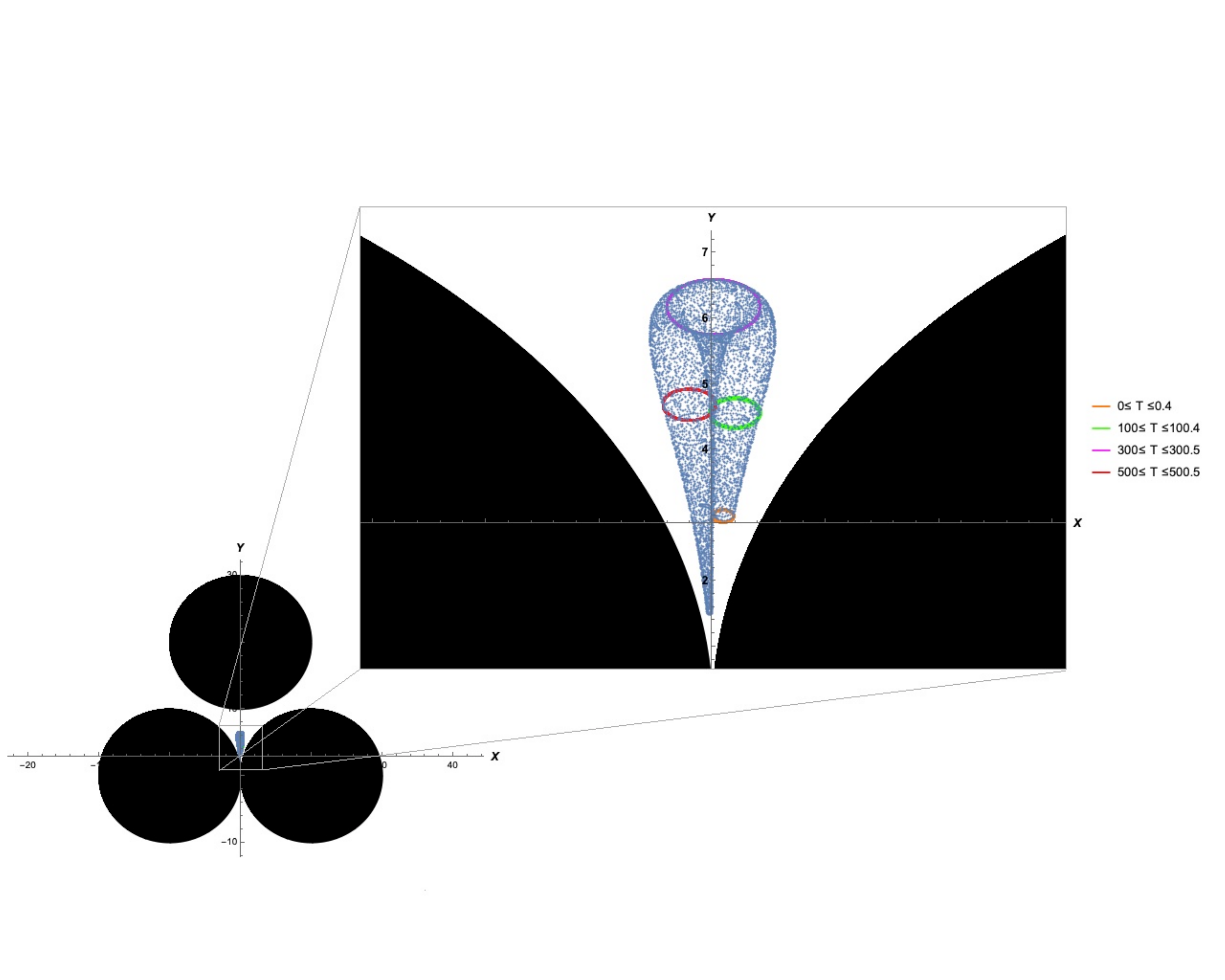}
    \caption{In this plot, we can see how the geodesic is trapped in the interior region formed by the three disks. Segments of trajectory associated to very short periods of time evolution have been marked with different colors. The blue dots are the positions of the particle sampled at long intervals.}
    \label{fig:Trapped1}
\end{figure}

The evaluated trajectory is shown in Figure \ref{fig:Trapped1}. We see that the geodesic gets effectively trapped. In order to clarify how the motion looks for small time periods, we have highlighted short intervals with separate colors.  These oscillations look like a particle in the presence of a magnetic field, and it is clear that these motions are repelled from the black disks. More precisely, the characteristic oscillation period of the trapped geodesics is very small with respect to the total time of numerical evolution $T$. Given the numerical evolution $T$, the geodesic has many oscillations that look as if they are filling the interior region. These show how the trajectory appears to be drifting slowly between different, almost integrable magnetically trapped orbits.
The total time evolution of the geodesic is represented by the blue color in Figure  \ref{fig:Trapped1}. 

We remark that the plot does not have the same scale in the $x$ and $y$ axes in order to better display the orbit of the geodesics. Because of this, the black regions that represent the locations of the disks are distorted. 
The Lyapunov exponents of this trajectory are all small and in our tests are consistent with values of zero. See the Appendix \ref{sec:small}. 

\begin{figure}[ht]
    \centering
    \includegraphics[width=8cm]{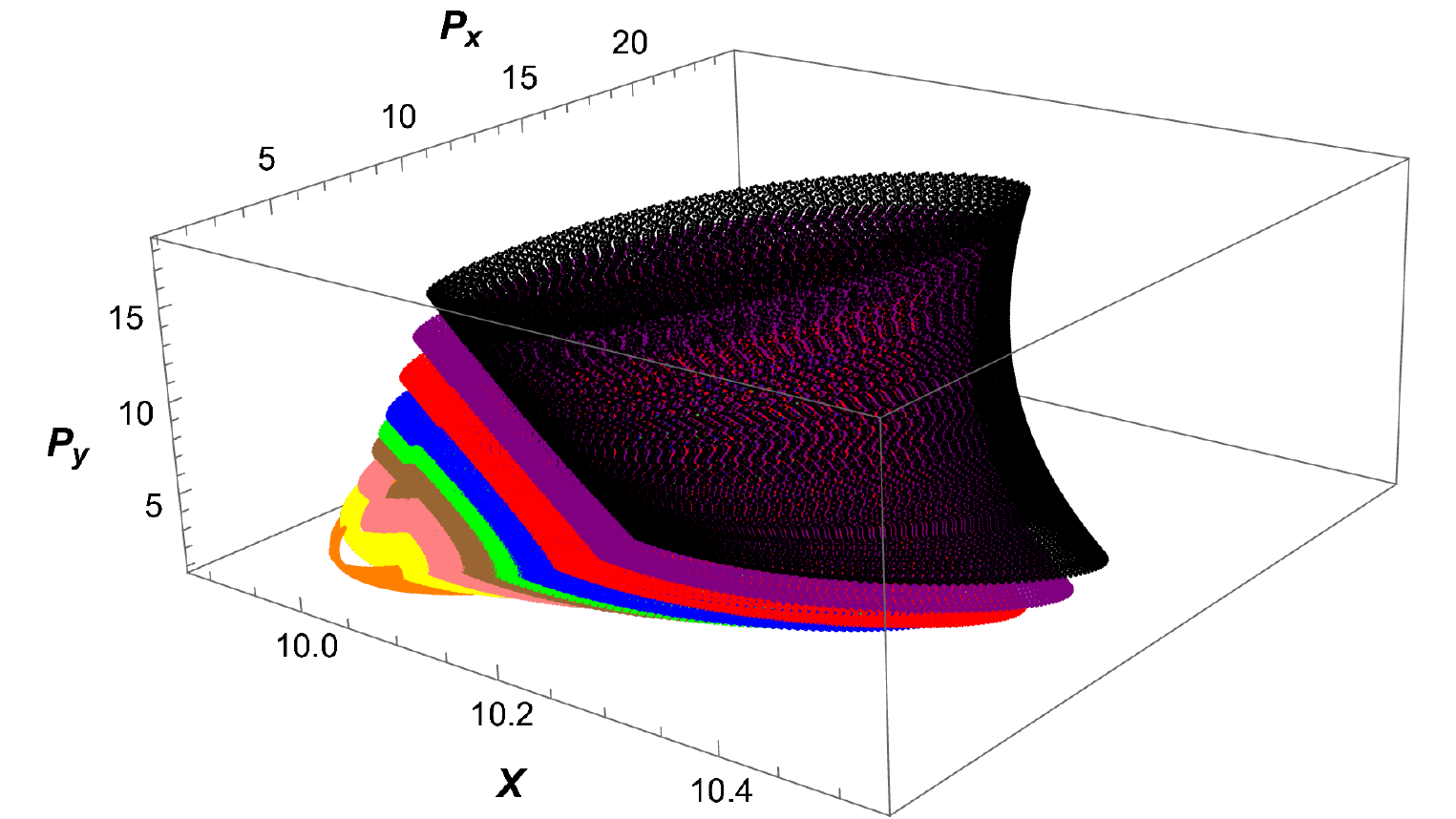}
    \caption{3d projection of the phase space  $x,P_x,P_y$. }
    \label{fig:PStrappedx}
\end{figure}

\begin{figure}[ht]
    \centering
    \includegraphics[width=8cm]{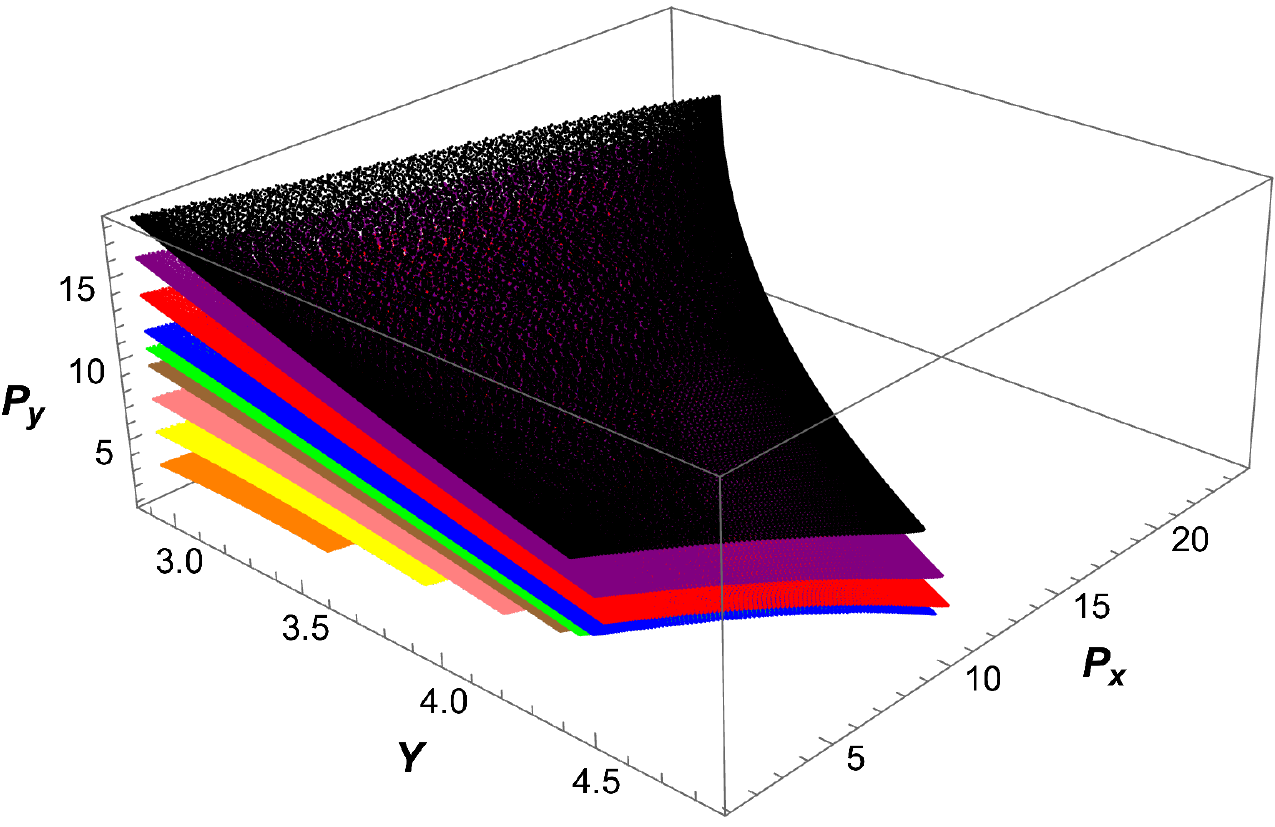}
    \caption{3d projection of the phase space  $y,P_x,P_y$. }
    \label{fig:PStrappedy}
\end{figure}

To more easily see that this orbit is actually chaotic,   in  Figures \ref{fig:PStrappedx} and \ref{fig:PStrappedy}, we present three-dimensional projections of the dynamics where we can see that the system is slowly filling the available phase space.
 This is accomplished by coloring different portions of the evolution in different colors and superposing them on top of each other. If the system were integrable, the trajectory would only fill a 2-torus, rather than a 3d volume. This would be readily apparent.
The long trajectory has not had enough time to fill its available phase space in \ref{fig:Trapped1}.
 Because of this, our Lyapunov exponent computation does not make complete sense: in the numerical limit defining the Lyapunov exponents, one assumes that the orbit has approximately explored all of its available phase space before one believes that the numerical value of the exponent has converged.

In these 3d plots, we see that the null geodesic is filling out the phase space. Hence, we conclude that the system is chaotic, but with a small Lyapunov exponent (we deal with this in the Appendix \ref{sec:small}), as the system seems to be quasi-adiabatically exploring the available phase space. 

\section{Conclusion}

In this paper, we have studied a special set of null geodesics in the general LLM geometries. These null geodesics are restricted to motion in the LLM plane. The reason that such geodesics stay in the LLM plane is that the LLM plane is a degeneration locus of symmetry (one of the two 3-spheres goes to zero radius) and this fact guarantees that solutions that preserve the symmetry must stay confined to the LLM plane.
 Because we are at a degeneration locus, the angular momentum associated to the shrunken sphere vanishes. The null geodesics are allowed to carry an angular momentum $L$ with respect to the remaining $SO(4)$. In our case, we have focused on situations where the angular momentum that is turned on is associated to the spatial symmetry of the boundary $S^3$, rather than the R-charge. 
 After the conserved quantities are taken into account, the effective motion is in two-dimensions and the corresponding phase space is four-dimensional. The motion is restricted to the region where the R-symmetry sphere is of zero size (the white region in our black and white coloring of the LLM geometries).
There are some special solutions of LLM geometries that preserve an additional $U(1)$ symmetry of rotations of the LLM plane. In these cases, the null geodesics restricted to the LLM plane are integrable.

When we studied particular cases where the rotational symmetry is broken, we found that the null geodesics exhibited chaotic behavior. Also, in general, we found that solutions that minimize the LLM energy after fixing the angular momentum satisfy the constraint $E=L$, stay at a fixed position in the LLM plane, and can be located arbitrarily on the allowed locus of the LLM plane (the white region). These are associated with states that satisfy a BPS constraint.
 This degeneration is similar to that of a classical particle in the lowest Landau level in a magnetic field. When we increase the energy, the orbit precesses. In some situations, the orbit's center moves slowly in the LLM plane, as depicted in Figure \ref{fig:Trapped1}, where $(E-L)/E$ is small. 

Let us consider what are the possible consequences of our observations. First, null geodesics are a geometric optics limit of massless wave equations on the LLM geometry.  We expect that there should a large class of BPS modes of supergravity fluctuations associated with these geodesics
that do not move. The fact that they are classically soluble suggests that they may also be found analytically in the ten-dimensional geometry.

Secondly, the fact that these in-plane geodesics are chaotic is indicative that the non-BPS protected modes in the LLM geometry can only be solved numerically. The modes associated with chaotic motion should display level statistics typical of chaotic systems. It becomes interesting to ask to what extent they display the eigenstate thermalization hypothesis properties. Solving these issues can provide interesting thermalization dynamics on these geometries.

More generally, we explored to what extent a generic LLM geometry might start looking like an incipient black hole. The interesting observation we have made is that the problem of solving these null geodesics becomes similar to a problem of chaotic billiards scattering in two dimensions. We expect that as we add more droplets the classical problem of {\em getting out of the black hole region} should resemble a problem of diffusive dynamics. 
Considering that we will eventually want to solve wave equations in these systems, one should try to understand to what extent other phenomena like Anderson localization due to the effective random potential might make certain modes much harder to probe than others.

These questions, once all symmetries are taken into account, should correspond to an eigenvalue problem of partial differential equations in three dimensions (the LLM plane variables $x,y$ plus the extra $\xi$ direction) that seem more tractable than other similar problems 
associated with fuzzball geometries, \cite{Cardoso:2008bp,Bena:2017upb,Bianchi:2017sds,Bianchi:2018kzy,Bianchi:2020des,Bianchi:2020yzr,Bacchini:2021fig,Addazi:2021pty,Heidmann:2022ehn,Emelin:2023mwy}. Moreover, it would be interesting to study more elaborated frameworks, for instance the simplest nontrivial LLM geometry: the annulus, or configurations that involve geodesics outside of the LLM plane. We hope to tackle some of these problems in the near future.

\acknowledgments
D.B. research supported in part by the Department of Energy under Award No DE-SC 0011702. R.M. research is supported in part by the Department of Energy under Award No. DE-SC0019139. AR is grateful for the hospitality of the Kavli Institute for Theoretical Physics, where this work was started. AR is partially supported by the INFN grant “Gauge Theories, Strings and Supergravity” (GSS). AR was also supported in part by the Heising-Simons Foundation, the Simons Foundation, and the National Science Foundation Grant No. NSF PHY-1748958.

\appendix

\section{Small Lyapunov exponents}\label{sec:small}

In this Appendix, we study numerically the effectively trapped geodesic discussed in section \ref{sec:trapped}.
Our goal is to measure the Lyapunov exponents. The maximum Lyapunov exponent is measured by taking an infinitesimal deformation $\delta$ in phase space and checking how quickly it grows.
The Lyapunov exponent is defined as
\begin{equation}
\lambda_{max} = \lim_{T\to \infty} \frac{1}{T}\log( |\delta(T)|/|\delta(0)|).
\end{equation}
The natural question to ask is if the total time $T$ is long enough to see that the maximum Lyapunov exponent is different from zero.

\begin{figure}[ht]
    \centering
    \includegraphics[width=10cm]{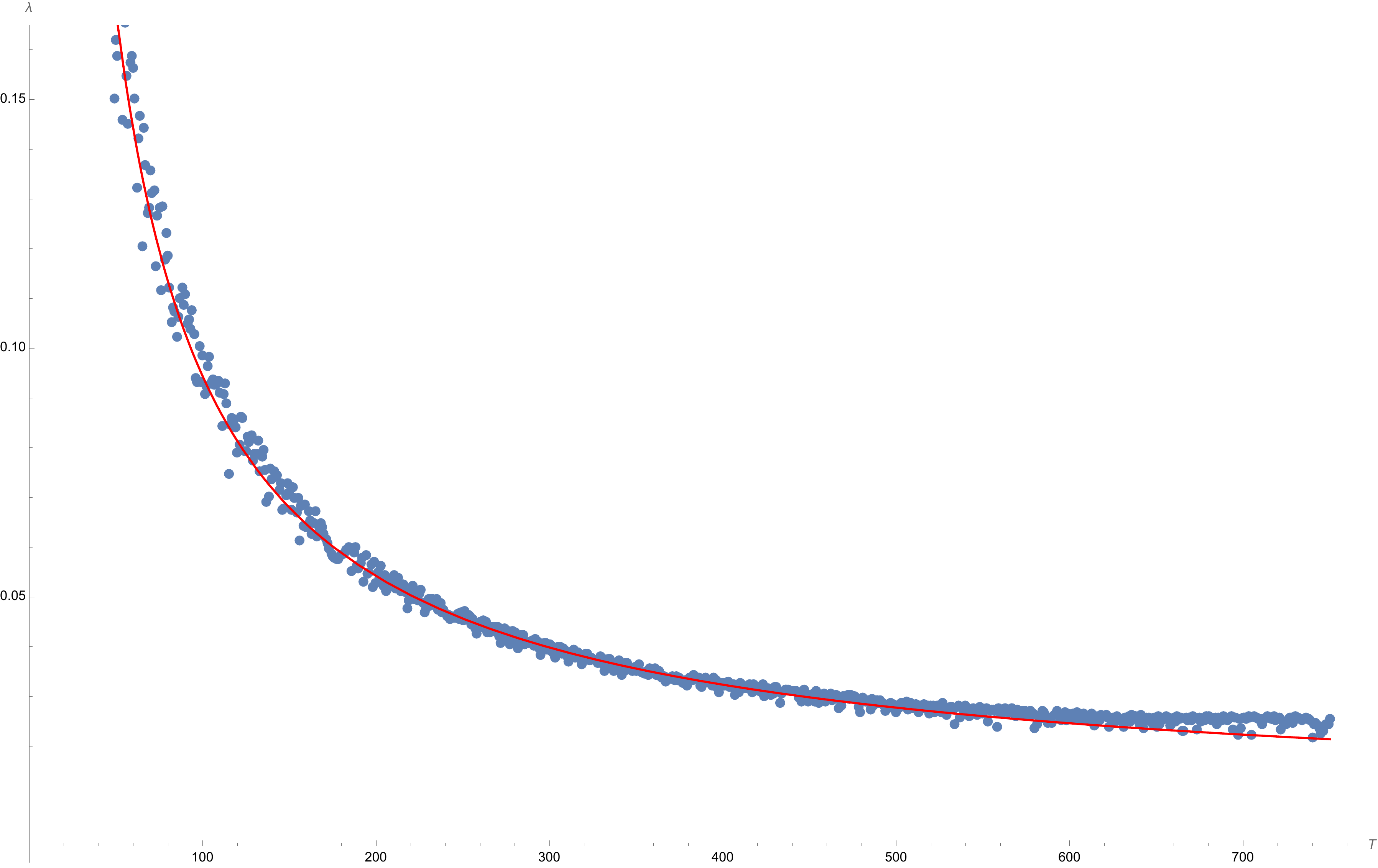}
    \caption{We plot the logarithm $\log( |\delta(T)|/|\delta(0)|)/T$ with respect to time.
    We show a fit of the corresponding quantity given by $0.00707+4.1227/T+\log(T)/T$ where $\lambda_{max}\simeq0.00707$ is interpreted as the maximum Lyapunov exponent. The other parts of the fit serve to correct for the quasi-adiabatic almost integrable behavior for short time scales.}
    \label{fig:LyapunovExpTrapped}
\end{figure}

In an integrable system, the growth of the maximal stretching direction is linear in time $T$ (the angle variables evolve linearly). 
The growth would therefore behave for large times as $\log(aT)$ and the Lyapunov exponent would only vanish in the $T\to \infty $ limit, decaying as $\log(T)/T$. We contrast the system we have with an integrable system by testing with a fit to $\log(T)/T$ in Figure \ref{fig:LyapunovExpTrapped}.

 It is clear that since the fit to an integrable system is fairly good, we cannot yet distinguish the system from an integrable system: the Lyapunov exponents have not converged yet. The infinite time limit of the fit would go to zero and
this is indicated by the maximum numerically evaluated Lyapunov exponent decaying slowly due to the logarithmic contribution. The fit gets worse at later times, indicating that there is a small Lyapunov exponent that will remain nonzero in the infinite time limit. However, this exponent is much smaller than what we were able to compute with this extended trajectory.

\end{document}